\documentclass{article}
\usepackage{amsmath}
\usepackage{appendix}
\usepackage[retainorgcmds]{IEEEtrantools}
\usepackage{graphicx}
\begin{document}
\title{Exact diagonalization of the truncated Bogoliubov Hamiltonian}
\author{Loris Ferrari \\ Department of Physics and Astronomy of the University (DIFA) \\Via Irnerio, 46,40127, Bologna, Italy}
\maketitle
\begin{abstract}
The present short note is simply intended to communicate that I have analytically diagonalized the Bogoliubov truncated Hamiltonian $H_c$~\cite{Bogo1,Bogo2}, in an interacting bosonic gas. This is the natural prosecution of my work~\cite{MS}, now denoted as (I), where the diagonalization was performed only in the subspace corresponding to zero momentum collective excitations (CE).\newline
\end{abstract}

e-mail: loris.ferrari@unibo.it
telephone: ++39-051-2095109\\

The Hamiltonian under consideration is:
\begin{align}
H_c&=\overbrace{\frac{\widehat{u}(0)N^2}{2}}^{E_{in}}+\sum_{\mathbf{k}\ne0}\overbrace{\left[\mathcal{T}(k)+\widetilde{N}_{in}\:\widehat{u}(k)\right]}^{\widetilde{\epsilon}_1(k)}b^\dagger_{\mathbf{k}}b_{\mathbf{k}}+\nonumber\\
\nonumber\\
&+\frac{1}{2}\sum_{\mathbf{k}\ne0}\widehat{u}(k)\Big[b^\dagger_{\mathbf{k}}b^\dagger_{-\mathbf{k}}(\:b_{\mathbf{0}}\:)^2+b_{\mathbf{k}}b_{-\mathbf{k}}(\:b^\dagger_{\mathbf{0}}\:)^2\Big]\:,\label{Hc}
\end{align}
\\ 
where $b^\dagger_{\mathbf{k}}$ and $b_{\mathbf{k}}$ create and destroy a spinless boson in the free-particle state $\langle\:\mathbf{r}\:|\:\mathbf{k}\:\rangle= e^{i\mathbf{k}\:\mathbf{r}}/\sqrt{V}$
and  

\begin{equation*}
\widehat{u}(q)=\frac{1}{V}\int\mathrm{d}\mathbf{r}e^{-i\mathbf{q}\:\mathbf{r}}\:u(r)\:,
\end{equation*}
\\
is the Fourier transform of the \emph{repulsive} interaction energy $u(r)$ ($>0)$. $\mathcal{T}(k)=\hbar^2k^2/(2M)$ is the kinetic energy. The number operator $\widetilde{N}_{in}=b^\dagger_{\mathbf{0}}b_{\mathbf{0}}$ refers to the bosons in the ground state, while $\widetilde{N}_{out}=\sum_{\mathbf{k}\neq\mathbf{o}}b^\dagger_{\mathbf{k}}b_{\mathbf{k}}$ will be used to numerate bosons in the excited states. Overtilded symbols indicate operators, to avoid confusion with their (non overtilded) eigenvalues.

The base I shall use for diagonalizing $H_c$ is formed by Fock states:

\begin{equation}
\label{| j >eta}
|\:j,\:\mathbf{k}\:\rangle_\eta=\frac{(b_\mathbf{0}^\dagger)^{N-2j-\eta}}{\sqrt{(N-2j-\eta)!}}\frac{(b_\mathbf{k}^\dagger)^{j+\eta}(b_\mathbf{-k}^\dagger)^j}{\sqrt{j!(j+\eta)!}}|\emptyset\:\rangle\:,
\end{equation}
\\
with $j+\eta$ (real) bosons in $|\:\mathbf{k}\:\rangle$, $j$ bosons in $|\:-\mathbf{k}\:\rangle$ and $N-2j-\eta$ bosons in $|\:\mathbf{0}\:\rangle$, so that the total momentum is, manifestly, $\eta\hbar\mathbf{k}$. 

Following the method developed in (I), I take advantage of the dependence on $k=|\:\mathbf{k}\:|$ of $\mathcal{T}(k)$ and $\widehat{u}(k)$, to express the Hamiltonian $H_c$ (eq.n~\eqref{Hc}) as a sum of independent one-momentum Hamiltonians

\begin{subequations}
\begin{equation}
\label{Hc2}
H_{c}=E_{in}+\sum_{\mathbf{k}\ne0}\widetilde{h}_c(\mathbf{k})\:,
\end{equation}
\\
where:

\begin{align}
\widetilde{h}_c(\mathbf{k})&=\frac{1}{2}\widetilde{\epsilon}_1(k)[b^\dagger_{\mathbf{k}}b_{\mathbf{k}}+b^\dagger_{-\mathbf{k}}b_{-\mathbf{k}}]+\nonumber\\
\nonumber\\
&+\frac{1}{2}\widehat{u}(k)\Big[b^\dagger_{\mathbf{k}}b^\dagger_{-\mathbf{k}}(\:b_{\mathbf{0}}\:)^2+b_{\mathbf{k}}b_{-\mathbf{k}}(\:b^\dagger_{\mathbf{0}}\:)^2\Big]\:.\label{DiagHceta}
\end{align}
\end{subequations}
\\
I study the \emph{exact} eigenstates of $\widetilde{h}_{c}(\mathbf{k})$ as linear combinations of the states eq.n~\eqref{| j >eta}:

\begin{subequations}
\begin{equation}
\label{| S >eta}
|S,\:\mathbf{k},\:\eta\:\rangle=\sum_{j=0}^\infty\phi_S(j,\eta)|j,\:\mathbf{k}\:\rangle_\eta\:,
\end{equation}
\\
by solving the eigenvalue equation

\begin{equation}
\widetilde{h}_c|S,\:\mathbf{k},\:\eta\:\rangle=\mathcal{E}_S(k,\eta)|S,\:\mathbf{k},\:\eta\:\rangle
\end{equation}
\end{subequations}
\\
in the unknowns $\phi_S(j,\eta)$, with boundary conditions $lim_{j\rightarrow\infty}\phi_S(j,\eta)=0$ (necessary for normalizability) and $\phi_S(-1,\eta)=0$ (exclusion of negative populations)\footnote{The reason why the sum in eq.n~\eqref{| S >eta} can be extended to $\infty$, in the TL, though $j$ cannot exceed the value $(N-\eta)/2$ is explained in detail in (I).}. It should be clear that $\eta$ can be assumed non negative, without loss of generality, and represents the asymmetry in the excited states populations with opposite momentum.

I provisionally drop the dependence on $\mathbf{k}$, $S$, $\eta$ , and express the energies in units of $N\widehat{u}(k)$: 

\begin{equation}
\label{underline}
\underline{\epsilon}_1=\frac{\epsilon_1(k)}{N\widehat{u}(k)},\:\underline{\mathcal{E}}=\frac{\mathcal{E}_S(k,\eta)}{N\widehat{u}(k)},\:\underline{\epsilon}=\frac{\epsilon(k)}{N\widehat{u}(k)}\:.
\end{equation}
\\
Then a slight generalization of the procedure adopted in Section 3 of (I) for equations~\eqref{| S >eta} and \eqref{DiagHceta} yields, in the TL:

\begin{align}
&\left[\underline{\epsilon}_1(2j+\eta)-2\underline{\mathcal{E}}\right]\phi(j)+\nonumber\\
\nonumber\\
&+\sqrt{(j+\eta)j}\:\phi(j-1)+\sqrt{(j+1+\eta)(j+1)}\:\phi(j+1)=0\:.\label{eq.phi}
\end{align}
\\
Thanks to the transformation:

\begin{equation}
\label{phi(j)}
\phi(j)=x^j\sqrt{\frac{(j+\eta)!}{j!\:\eta!}}P(j)\:,
\end{equation}
\\
equation~\eqref{eq.phi} yields, in turn:

\begin{align}
&\left[\underline{\epsilon}_1(2j+\eta)-2\underline{\mathcal{E}}\right]P(j)+\nonumber\\
\nonumber\\
&+\frac{j}{x}P(j-1)+x(j+1+\eta)P(j+1)=0\:,\label{eq.P}
\end{align}
\\
in the new unknowns $x$ and $P(j)$. As shown in Section 3 of (I), a finite difference equation in the form \eqref{eq.P} can be solved by assuming for $P(j)$ a $S$-degree polinomial expression:

\begin{equation}
\label{P}
P(j)=\sum_{n=0}^SC(n)\:j^n\:.
\end{equation}
\\
A system of $S+2$ linear equations for the unknowns $C(j)$ then follows from the vanishing of each term proportional to $j^n$, with $n=0,\:1,\;\dots,\:S+1$:

\begin{align}
&\left(\underline{\epsilon}_1\eta-2\underline{\mathcal{E}}\right)C(n)+ 2\underline{\epsilon}_1C(n-1)+\nonumber\\
\nonumber\\
&+x\left[(1+\eta)\sum_{s=n}^SC(s)\binom{n}{s}+\sum_{s=n-1}^SC(s)\binom{n}{s-1}\right]-\nonumber\\
\nonumber\\
&-\frac{1}{x}\sum_{s=n-1}^SC(n)\binom{n}{s-1}(-1)^{n-s+1}=0\:,\label{eq.C}
\end{align}
\\
($C(S+1)=0$ by definition). I am especially interested in the two highest order equations ($n=S+1,\:S$) and in the lowest order one ($n=0$). From the system~\eqref{eq.C} one gets:

\begin{subequations}
\begin{align}
&C(S)\left[2\underline{\epsilon}_1+x+x^{-1}\right]=0&\quad(n=S+1)\label{S+1}\\
\nonumber\\
&C(S-1)\left[2\underline{\epsilon}_1+x+x^{-1}\right]+\nonumber\\
\nonumber\\
&+C(S)\left[\underline{\epsilon}_1\eta-2\underline{\mathcal{E}}+S\left(x-x^{-1}\right)+x(1+\eta)\right]=0&\quad(n=S)\label{S}\\
\nonumber\\
&\dots
\nonumber\\
&\left(\underline{\epsilon}_1\eta-2\underline{\mathcal{E}}\right)C(0)+x(1+\eta)\sum_{s=0}^SC(s)&\quad(n=0)\:.\label{0}
\end{align}
\end{subequations}
\\
For any non vanishing value of $C(S)$, equation \eqref{S+1} determines $x$, according to the equation $x^2+2x\underline{\epsilon}_1+1=0$, with the condition $|x|<1$ (normalizability):

\begin{equation}
\label{x}
x=\underline{\epsilon}-\underline{\epsilon}_1\quad;\quad x^{-1}=-\left(\underline{\epsilon}+\underline{\epsilon}_1\right)
\end{equation}
\\
with

\begin{align}
\label{epsilon}
\epsilon(k)&=\sqrt{\mathcal{T}^2(k)+2N\:\widehat{u}(k)\mathcal{T}(k)}=\nonumber\\
\nonumber\\
&=\frac{\hbar\:k}{\sqrt{2M}}\sqrt{2N\:\widehat{u}(k)+\frac{\hbar^2k^2}{2M}}=\nonumber\\
\nonumber\\
&=N\widehat{u}(k)\sqrt{\underline{\epsilon}_1^2-1}\:.
\end{align}
\\
(remember eq.ns \eqref{underline}). Then, equation~\eqref{S} determines $\underline{\mathcal{E}}$, according to the equation:

\begin{equation}
\underline{\epsilon}_1\eta-2\underline{\mathcal{E}}+S\left(x-x^{-1}\right)+x(1+\eta)=0\:,
\end{equation}
\\
that, with the aid of eq.n~\eqref{x} and with $k$, $S$, $\eta$, $N\widehat{u}(k)$ restored, yields the complete energy eigenvalues:

\begin{align}
\mathcal{E}_S(k,\:\eta)&=\frac{\epsilon(k)}{2}\left(\eta+\frac{1}{2}\right)+\epsilon(k)\left(S+\frac{1}{2}\right)-\label{E(S,eta)}\\
\nonumber\\
&-\underbrace{\left[\frac{\epsilon_1(k)}{2}+\frac{3\epsilon(k)}{4}\right]}_{\mathcal{E}_0(k)}\quad(S,\:\eta=0,\:1,\:\dots)\nonumber\:,
\end{align}
\\

The algebraic structure of the eigenvalue problem~\eqref{eq.C} is fairly peculiar: equations~\eqref{S+1} and \eqref{S} determine $x$ and $\underline{\mathcal{E}}$ (i.e. the exponential decay in $j$ and the energy eigenvalue). The still arbitrary coefficients $C(S)$ and $C(S-1)$ enter the next $S$ equations, that can be solved for the $S$ unknowns $C(1),\:C(2),\dots,\:C(N)$ in terms of, say, $C(0)$, that will be determined by normalization.

Finally, recalling eq.ns~\eqref{P} and \eqref{phi(j)}, it is easy to see that equation \eqref{0} coincides with the boundary condition following from eq.n~\eqref{eq.phi}\footnote{The boundary condition at $j=-1$ plays an important role in selecting the eigensolution. Actually, a different choice $\phi(j)=x^j\sqrt{j!\:\eta!/(j+\eta)!}P(j)$ would yield a $S$-degree polinominial solution for $P(j)$, just like eq.n~\eqref{phi(j)}, but the boundary condition $\phi(-1,\eta)=0$ would not be satisfied.}:

\begin{equation}
\phi(-1)=0\:\Rightarrow\:\left(\underline{\epsilon}_1\eta-2\underline{\mathcal{E}}\right)\phi(0)+\phi(1)\sqrt{1+\eta}=0\:.
\end{equation}
\\
Restoring all the relevant entries determining the state~\eqref{| S >eta}, and recalling eq.ns~\eqref{phi(j)}, \eqref{P}, one finally gets:

\begin{equation}
\label{|S,eta>2}
|S,\:\mathbf{k},\:\eta\:\rangle=\sum_{j=0}^\infty \overbrace{x^j\sqrt{\binom{j+\eta}{j}}\sum_{s=0}^SC_S(s,\eta)j^n}^{\phi_S(j,\eta)}|j,\:\mathbf{k}\:\rangle_\eta\:.
\end{equation}
\\
The probability amplitude on the Fock states $|j,\:\mathbf{k}\:\rangle_\eta$ (eq.n~\eqref{| j >eta}) reads, asymptotically:

\begin{equation}
\left|\phi_S(j,\eta)\right|^2\rightarrow j^{2S+\eta}x^{2j}\:,
\end{equation}  
\\
showing that the total number $2S+\eta$ of energy quanta $\epsilon(k)/2$ (see eq.n~\eqref{E(S,eta)}), associated to an eigenstate eq.n~\eqref{|S,eta>2}, coincides with the highest power of $j$, in the limit $j>>1$. 

Recalling that the eigenvalues \eqref{E(S,eta)} must be counted \emph{twice} in the sum \eqref{Hc2} \footnote{This point escaped from my attention in ref. \cite{MS}. A corrigendum has been sent to Physica B, to emend the error.}, the main results of the exact diagonalization of $H_c$ are:\newline
\\
(A) The energy spectrum results from the sum of \emph{two} independent oscillators, one with frequency $2\epsilon(k)/h$, labeled by the non negative integer $S$; the other with halved frequency $\epsilon(k)/(h)$, labeled by the non negative integer $\eta$.
\\
(B) If, as reasonable, the quasiphonons's must behave like (massless) particles and, thereby, must carry a finite momentum, the generating oscillator is the one labeled  by $\eta$ and the energy of each quasiphonons is $\epsilon(k)$.      
\\

The physical consequences of these results and a comparison with the current picture of Bogoliubov/Landau CE's in the interacting boson gas will appear in a forthcoming paper.

\end{document}